\documentclass[preprint,prd,aps,tighten,nofootinbib,amssymb]{revtex4}

\usepackage[dvipdfmx]{graphicx}
\usepackage{epsfig,amssymb,amsmath}
\usepackage{subfigure}
\usepackage[colorlinks=true,linkcolor=blue,citecolor=blue]{hyperref}
\usepackage{float}
\usepackage{here}

\makeatletter
\def\Hline{%
\noalign{\ifnum0=`}\fi\hrule \@height 2pt \futurelet
\reserved@a\@xhline}
\makeatother

\def\a{\alpha}
\def\b{\beta}

\def\h{\theta}
\def\k{\kappa}
\def\l{\lambda}
\def\m{\mu}
\def\n{\nu}

\def\r{\rho}

\def\D{\Delta}
\def\G{\Gamma}

\def\L{\Lambda}

\newcommand{\beq}{\begin{equation}}
\newcommand{\eeq}{\end{equation}}
\newcommand{\bea}{\begin{eqnarray}}
\newcommand{\eea}{\end{eqnarray}}
\newcommand{\bear}{\begin{array}}
\newcommand {\eear}{\end{array}}
\newcommand{\bef}{\begin{figure}}
\newcommand {\eef}{\end{figure}}
\newcommand{\bec}{\begin{center}}
\newcommand {\eec}{\end{center}}
\newcommand{\non}{\nonumber}

\def\GEV#1{10^{#1}{\rm\,GeV}}

\def\lrf#1#2{ \left(\frac{#1}{#2}\right)}
\def\lrfp#1#2#3{ \left(\frac{#1}{#2} \right)^{#3}}

\begin{document}
\draft
\tighten
\preprint{TU-1031,~IPMU16-0119,~SISSA 45/2016/FISI}
\title{\large \bf Dark Matter in  Axion Landscape
}
\author{
    Ryuji Daido\,$^{a}$\footnote{email: daido@tuhep.phys.tohoku.ac.jp},
    Takeshi Kobayashi\,$^{b,c}$\footnote{email: takeshi.kobayashi@sissa.it},
    Fuminobu  Takahashi\,$^{a,d}$\footnote{email: fumi@tuhep.phys.tohoku.ac.jp}
    }
\affiliation{
$^a$ Department of Physics, Tohoku University, Sendai 980-8578, Japan,\\
$^b$ SISSA, Via Bonomea 265, 34136 Trieste, Italy,\\
$^c$ INFN, Sezione di Trieste, Via Bonomea 265, 34136 Trieste, Italy, \\
$^d$ Kavli IPMU, TODIAS, University of Tokyo, Kashiwa 277-8583, Japan
}

\vspace{2cm}

\begin{abstract}
If there are a plethora of axions in nature, they may have a complicated potential and 
create an {\it axion landscape}.
We study a possibility that one of the axions is so light that it is cosmologically stable,
explaining the observed dark matter density. In particular we focus on a case in which
two (or more) shift-symmetry breaking terms conspire to make
the axion sufficiently light at the potential minimum.
 In this case the axion has a flat-bottomed potential.
In contrast to the case in which a single cosine term dominates the potential, 
the axion abundance as well as its isocurvature perturbations are significantly suppressed.
This allows an axion with a rather large mass to serve as dark matter without fine-tuning of the
initial misalignment, and
further makes higher-scale inflation to be consistent with the scenario.
\end{abstract}

\maketitle

\section{Introduction}
\label{intro}
In string theory, there appear many  axions in the compactification of the extra 
dimensions~\cite{Witten:1984dg,Green:1987mn}. The axions typically respect discrete shift symmetries and have
periodic potentials with decay constants of order $f\sim10^{15-16}\,{\rm GeV}$~\cite{Svrcek:2006yi}. 
The Universe with a plenitude of those axions whose masses range over many orders
of magnitude is called the Axiverse~\cite{Arvanitaki:2009fg}, and  its phenomenological and
cosmological implications have been extensively studied in the 
literature (e.g. \cite{Arvanitaki:2010sy,Cicoli:2012sz,Marsh:2011bf,Daido:2015bva}).

The axion potential arises from various non-perturbative effects, and 
the axions generically have both kinetic and mass mixings. If there are a plethora of axions, their potential 
may well be quite complicated. Then, the axions create an
{\it axion landscape} with many local minima and maxima~\cite{Higaki:2014mwa,Higaki:2014pja}, where 
eternal inflation as well as quantum tunneling/jump-up between adjacent local minima 
take place repeatedly. The axion landscape may be a low-energy branch of the
string landscape~\cite{Susskind:2003kw}.

The axions are known to be a plausible candidate for the inflaton.
If multiple axions have an aligned potential~\cite{Kim:2004rp,Choi:2014rja}, 
the effective decay constant of the lightest axion can be super-Planckian. Then slow-roll inflation naturally 
takes place along the direction after tunneling from adjacent vacua~\cite{Higaki:2014mwa,Higaki:2014pja}.
In this case, the inflaton potential generically has small modulations and may generate a sizable
running of the spectral index~\cite{Kobayashi:2010pz,Czerny:2014wua}. It is also possible 
to realize an axion hilltop inflation with sub-Planckian decay constants if two (or more) cosine terms conspire to
generate a flat-top potential. Such inflation with multiple cosine potentials is called
multi-natural inflation~\cite{Czerny:2014wza,Czerny:2014xja,Czerny:2014qqa}, and
in contrast to  natural inflation with a single cosine term~\cite{Freese:1990ni,Adams:1993ni},
there is no lower bound on the decay constant. See also Refs.~\cite{Croon:2014dma,Higaki:2015kta,Kadota:2016jlw} for the works along a similar line.

In this paper we study a possibility that  the axions play another important 
cosmological role, i.e., dark matter.\footnote{
The axion may also be responsible for the origin of baryon asymmetry~\cite{Kusenko:2014uta,Daido:2015gqa}.
} If there are many axions as suggested by the Axiverse or axion landscape, 
one of the axions may be so light that it is cosmologically stable and contributes to dark matter~\cite{Preskill:1982cy,Abbott:1982af,Dine:1982ah}.
See e.g. Refs.~\cite{Kim:2008hd,Bae:2008ue,Wantz:2009it,Ringwald:2012hr,Kawasaki:2013ae,Marsh:2015xka}
 for recent reviews on the axion dark matter.
Broadly speaking, there are two ways to realize a sufficiently light axion mass. One is to suppress
all possible terms in the axion potential. This is the case if all the dynamical scales of the 
interactions responsible for the axion potential are sufficiently suppressed.
In this case the axion dynamics can be studied by approximating that 
 a single term dominates the potential,  as usually assumed in the literature. 
 The other is that two or more shift-symmetry breaking terms conspire to make the axion sufficiently light at the potential minimum.\footnote{Such a cancellation naturally takes place in toroidal compactifications where the moduli potential is given
 by the Jacobi theta functions~\cite{Higaki:2015kta}.
 }
Then, the axion has a flat-bottomed potential where the curvature at the potential minimum is suppressed compared to its typical value.
Such a potential is similar to the inverted potential of the axion hilltop inflation~\cite{Czerny:2014wza,Czerny:2014xja,Czerny:2014qqa}. 

The purpose of this paper is to investigate the axion dynamics and estimate the axion abundance and its isocurvature perturbations in the flat-bottomed potential realized by multiple shift-symmetry breaking terms.
As we shall see shortly, the axion abundance as well as its isocurvature perturbations can be
significantly suppressed compared to the case in which a single cosine
term dominates the potential.
This allows an axion with a rather large mass to serve as dark matter, and further makes higher-scale inflation to be consistent with the scenario.
We will  show  that the axion mass can be as heavy as $m\sim100$ MeV and such a relatively heavy axion may decay into hidden photons
with a lifetime close to the present age of the Universe, which partially relaxes the tension of $\sigma_8$~\cite{Battye:2014qga,Enqvist:2015ara}.\footnote{If the flat-bottomed potential emerges from the anthropic requirement for cosmologically stable dark matter, the bound
on the lifetime may be saturated.}

Let us mention related works in the past. The inflaton potential with modulations was extensively studied
in  the axion monodromy inflation~\cite{Silverstein:2008sg,McAllister:2008hb,Hannestad:2009yx}.
In addition to inflation, the potential with small modulations has been
studied in various context. For instance,
the axion potential with small modulations was studied in a context of a curvaton model~\cite{Takahashi:2013tj}, the aligned QCD axion~\cite{Higaki:2016yqk} and monodromy dark matter~\cite{Jaeckel:2016qjp}. In those works, the axion potential is given by cosine terms with relatively large hierarchy in the potential heights and periodicities. 
Such a large hierarchy is not necessarily required for our purpose.

The rest of this paper is organized as follows. In Sec.~\ref{axionlandscape}, 
we explain our set-up and give the axion dark matter models. In Sec.~\ref{axionabundance} and \ref{isocurvature}
we study the evolution of the axion and estimate the axion abundance and its isocurvature perturbations.
The last section is devoted to  conclusions.

\section{Axion Dark Matter}
\label{axionlandscape}
\subsection{Axion potentials}
In the axion landscape, the axions typically have the following potential:
\begin{align}
V(\phi_\alpha) = \sum_{i=1}^{N_S} \Lambda_i^4  
\left(1-\cos\left(\sum_{\alpha=1}^{N_A} n_{i\alpha} \frac{\phi_\alpha}{f_\alpha} + \delta_i \right)\right) + C,
\label{ALV}
\end{align}
where the prefactor $\Lambda_i$ and $\delta_i$ denote the dynamical scale and a CP phase of the corresponding 
non-perturbative effect, $n_{i \alpha}$ is an integer-valued anomaly coefficient matrix, $f_\alpha$ is the decay constant,  
and $N_S$ and $N_A$ are the
 number of shift symmetry breaking terms and axions, respectively. If $N_S > N_A$, there
appear many local minima, and the axion potential can be quite
complicated, forming an axion landscape~\cite{Higaki:2014mwa,Higaki:2014pja}.
The constant $C$ is introduced to set the cosmological constant vanishingly
small in our vacuum. In the early Universe, the axion may be trapped in one of the local minima where 
eternal inflation takes place. At a certain point, it tunnels through the potential barrier toward one of the adjacent local 
minima with a lower energy. If the axion potential there is sufficiently flat, slow-roll inflation may take over. The potential
can be flat either by the alignment or by the cancellation among shift-symmetry breaking terms.

In the axion landscape  the axion potential generically receives multiple contributions with
different heights and periodicities. Motivated by this observation, we consider a possibility that one of the axions is so light
that it is stable in a cosmological time scale and becomes a dominant component of dark matter.
To be concrete, we consider the following two models,
\beq
V(a)=\L^4\left[1-\cos\left(\frac{a}{f}\right)\right]
\label{single}
\eeq
and
\beq
V(a)=\L_1^4\left[1-\cos\left(n_1\frac{a}{f}\right)\right]+\L_2^4\left[1-\cos\left(n_2\frac{a}{f}+\delta\right)\right]+C,
\label{double}
\eeq
where we assume that the dynamical scales are constant in time
during and after inflation.
 For later use, we define  
\beq
M_i\equiv \frac{n_i\L_i^2}{f},~~~{\rm for~~}i=1,2,
\eeq
which represents the curvature scale of each term. We also define a dimensionless axion field, $\theta \equiv a/f$.

\subsection{Axion mass and lifetime}
\begin{figure}[t]
\centering
\includegraphics[width=9cm]{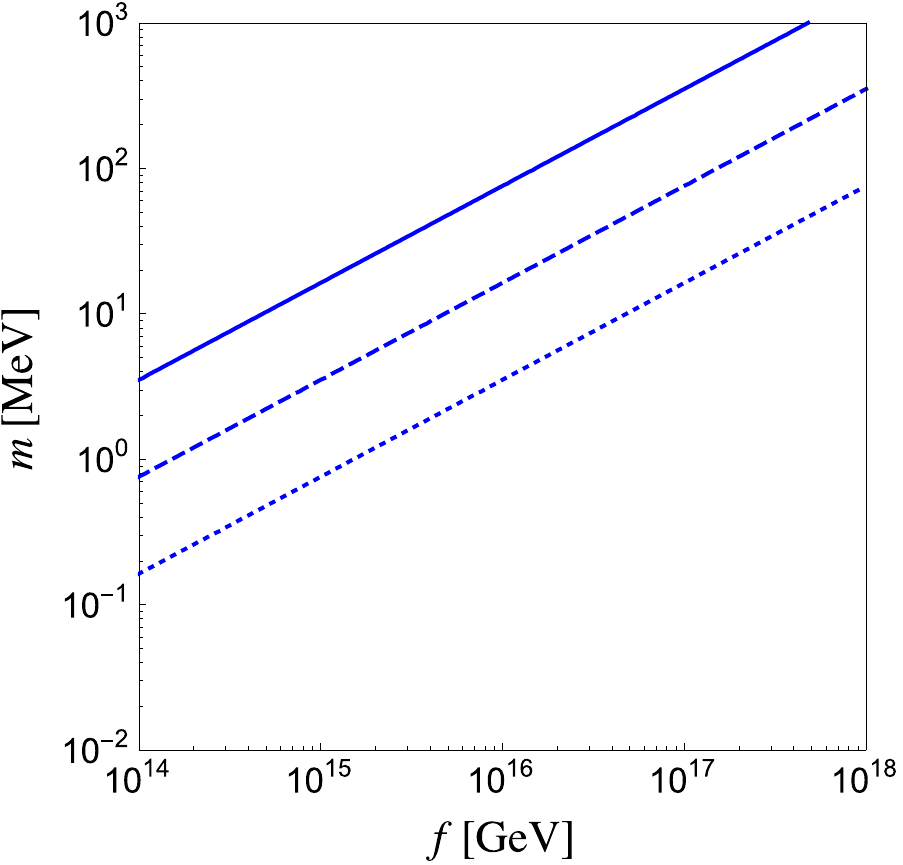}
\caption{Upper bound on axion mass from the lifetime constraint on dark matter in $f-m$ plane,
for different values of $\alpha_H = 0.01$ (top solid), $0.1$ (middle dashed), and $1$ (bottom dotted), respectively.
}
\label{life}
\end{figure}

The axion needs to be sufficiently long-lived to explain the observed dark matter density.
This requires the axion mass at the potential minimum,  $m^2\equiv V''(a_{\mathrm{min}})$, to be small enough. 
To get the feel of how light the axion should be, let us assume that
the axion is coupled to hidden photons as 
\beq
\frac{\alpha_H \,a}{4 \pi f} F_{H \m\n}\tilde{F}_H^{\m\n},
\eeq
where $F_H$ and $\tilde{F}_H$ are the field strength of the hidden photon field and its dual, respectively,
and $\alpha_H$ is the hidden fine-structure constant. 
The decay rate of the axion through the above operator is given by
\beq
\G_a=\frac{\a_H^2}{64\pi^3}\frac{m^3}{f^2}.
\eeq
The lifetime of dark matter is constrained to be longer than $97$ Gyr~\cite{Enqvist:2015ara},
which constrains the axion mass and decay constant as
\beq
m \;\lesssim\; 75 {\rm\,MeV} \lrfp{\alpha_H}{10^{-2}}{-\frac{2}{3}} \lrfp{f}{\GEV{16}}{\frac{2}{3}}.
\label{uppm}
\eeq
In Fig.~\ref{life} we show the constraint with $\a_H = 0.01, 0.1$ and $1$ from top to bottom. 
As one can see, for $f \simeq \GEV{16}$ and $\alpha_H = 0.01$, the upper bound on the
axion mass is about $100$ MeV. 
In particular, if the upper bound on the axion mass is saturated, the lifetime of the axion dark matter is not very far from the
present age of the Universe. In this case, the recent decrease in the dark matter density could reduce the growth of large-scale structure,
relaxing the tension between the observed value of $\sigma_8$ and the value inferred by the CMB observations
based on the $\Lambda$CDM model~\cite{Battye:2014qga,Enqvist:2015ara}.

\subsection{Flat-bottomed potential}
Roughly speaking, there are two possibilities to realize such a small axion mass. One is to suppress all the pre-factors of the terms relevant for the axion of our interest, and there is no special cancellation among different terms. In this case the axion potential can be approximated by Eq.~(\ref{single}),  assuming that a single cosine term dominates the potential.
Unless the initial position is close to the potential maximum,
the axion potential can be further simplified to a quadratic potential,
\beq
V(a) \;\simeq\; \frac{m^2}{2} a^2
\label{appquad}
\eeq
with $m \equiv \Lambda^2/f$.
The anharmonic effect due to the deviation from the quadratic term was studied 
in e.g. Refs.~\cite{Lyth:1991ub,Bae:2008ue,Kobayashi:2013nva}.

\begin{figure}[t]
\centering
\includegraphics[width=9cm]{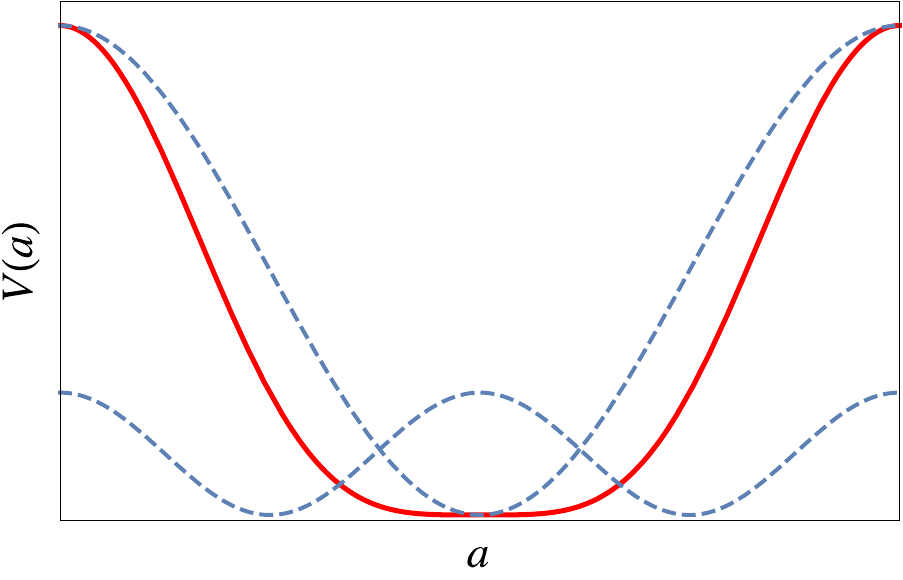}
\caption{
The schematic picture of the flat-bottomed potential (solid red) from two cosine terms
(dashed blue).
}
\label{potential}
\end{figure}

The other is to rely on cancellation among different terms.
  In the case of $\delta = \pi$, 
 the scalar potential (\ref{double}) can be 
 expanded about the potential minimum at the origin,
\beq
V(a) \;\simeq\; \frac{m^2}{2}a^2+\frac{\l}{4!} a^4,
\label{appquart}
\eeq
where $m$ and $\l$ are given by
\beq
m^2\equiv M_1^2-M_2^2
\eeq
and
\beq
\l\equiv\frac{n_2^2M_2^2-n_1^2M_1^2}{f^2},
\eeq
respectively. 
 Therefore, the axion mass $m$ is suppressed if $M_1^2$ is sufficiently close to $M_2^2$.
 In the following we assume $M_1^2 \simeq M_2^2 \gg m^2 > 0$
 and $n_2 > n_1 > 0$. 
 In this case,
 the quartic coupling can be approximated by
 \beq
 \lambda \simeq (n_2^2-n_1^2)\frac{M_1^2}{f^2}.
 \eeq
Higher order terms such as $\propto a^6$ are
suppressed compared to the quartic term
for $|a| \lesssim f$, given that $n_1$ and $n_2$ do not take huge values. 
In Fig.~\ref{potential} we show the schematic picture of the flat-bottomed axion potential.\footnote{
Note  that, if the signs of the quadratic and 
quartic couplings are assumed to be negative, the axion potential is exactly the same as
that for the axion hilltop inflation~\cite{Czerny:2014wza,Czerny:2014xja,Czerny:2014qqa}.}
If $\delta \ne \pi$, linear and cubic terms appear and the potential minimum is also shifted.
However, our analysis remains valid if $|\pi-\delta| \lesssim (m/M_1)^3$, in which case
the axion potential is still approximately given by Eq.~(\ref{appquart})
around the transition point $\h = \h_T$ defined below.

In both cases, the axion can be so light that it contributes to the observed dark matter density.
The axion abundance as well as its isocurvature perturbation, however, are quite different in the two cases,
because they crucially depend on the form of the axion potential.
This can be easily seen by noting that,
in the latter case, the axion potential is dominated by the quartic potential except for the vicinity of the origin.
We define a transition point, 
 \beq
\h_T \;\equiv\; \frac{m}{f}\sqrt{\frac{6}{\l}}\simeq \sqrt{\frac{6}{n_2^2-n_1^2}}\frac{m}{M_1} \ll 1,
\eeq
at which the quadratic and quartic terms give the equal contributions to the equation of motion for the axion.
Namely, the axion potential is approximated by the quartic (quadratic) term for $\theta \gtrsim  \theta_T$ ($\theta \lesssim \theta_T$).
The transition point $\h_T$ is much smaller than unity, and so, for a general initial condition, the initial misalignment
is much larger than $\theta_T$. Thus, the axion starts to oscillate in a quartic potential and its energy density
decreases like radiation, which significantly reduces the final axion abundance.
This lasts  until the oscillation amplitude becomes smaller than $\theta_T$, and then
the axion behaves like non-relativistic matter as usual. 
Furthermore, the isocurvature perturbation is also suppressed since the
quartic potential can wipe out the initial field fluctuations as the axion
field is forced to roll down along its
potential~\cite{Kobayashi:2012ba}.
This effect is particularly prominent if the curvature of the
potential is comparable to the Hubble parameter during inflation,
which can be the case with a quartic potential. 
This should be contrasted to the familiar case with a quadratic potential
where the curvature of the axion potential is negligibly small
everywhere along the potential, if
the axion has a sufficiently light mass  to explain dark matter.
In the following sections, we will study the axion dynamics in detail.

\section{Dark Matter Abundance}
\label{axionabundance}
In this section we study the evolution of the homogeneous mode of the axion field
to estimate its abundance.
To this end, we follow the axion dynamics from during inflation when
the CMB scales exit the horizon, and determine when in the
post-inflation epoch the axion starts to oscillate.
Specifically we will express the field value at the onset of oscillations, 
$\theta_{\rm osc}$, as a function of the field value at the horizon
exit, $\theta_*$.
After inflation the universe is considered to undergo an (effectively)
matter-dominated era, and then a radiation-dominated era after
reheating. The axion may start its oscillation before or after reheating,
depending on the reheat temperature.
Here and in what follows, the subscripts, `$*$', `osc', `reh',
represent that the variable is evaluated at the horizon exit, the onset of oscillations, and reheating, respectively.

We assume that 
the axion is deviated from the potential minimum during inflation and it starts to oscillate after inflation.
Before the onset of the oscillations, the axion dynamics  is described by an attractor solution,
whose equation of motion is given by~\cite{Kawasaki:2011pd}
\beq
cH\dot{a}\simeq-V',\label{attractor}
\eeq 
where { the dot  and prime denote a derivative with respect to time and axion field $a$, respectively, and} $c$ is a constant given by
\beq
c=
\begin{cases}
3 & (\text{de Sitter})\\
9/2 & (\text{matter dominant})\\
5 & (\text{radiation dominant})
\end{cases}.
\eeq
The above equation of motion for the attractor solution is valid if the curvature of the potential, $|V''|$, is much smaller
than the Hubble parameter squared (see Appendix A of Ref.~\cite{Kawasaki:2011pd} for the derivation). 
Note that $c$ is  no longer constant if $|V''|$ is larger than or comparable to the Hubble parameter.
Assuming that the axion dynamics follows the attractor solution, one can
integrate Eq.~(\ref{attractor}) from the horizon exit to the onset of  oscillations. 
Using $\dot{H}/H^2=3-c$,
 we obtain
\beq
\int^{a_{\rm osc}}_{a_*}\frac{da}{V'}\simeq-\frac{N_*}{3H_{\rm inf}^2}-\frac{1}{2c(c-3)H_{\rm osc}^2},\label{starosc}
\eeq
where $N_*$ is the number of e-folds during inflation between the horizon exit of
the CMB pivot scale with $k_p = 0.05\,~{\rm Mpc}^{-1}$ and the end of inflation.
The constant $c$ in this expression corresponds to the value right before the
onset of oscillations, i.e.,
$c = 9/2$ if $t_{\mathrm{osc}} < t_{\mathrm{reh}}$, and
$c = 5$ if $t_{\mathrm{osc}} > t_{\mathrm{reh}}$.
Upon obtaining the right hand side we have assumed that the axion starts
to oscillate well after 
inflation, $H_{\rm inf}^2\gg H_{\rm osc}^2$.
In the case where $t_{\mathrm{osc}} > t_{\mathrm{reh}}$, we further
assumed $H_{\rm reh}^2\gg H_{\rm osc}^2$.

We define the onset of the oscillation as
when the field variation during one Hubble time becomes comparable to
the distance to the potential minimum, i.e.,
\beq
\left|\frac{\dot{a}}{Ha}\right|_{\rm osc}=\k,\label{onset}
\eeq
where $\k$ is a constant of order unity, and we fix it to $\kappa=1/2$ hereafter. 
Thus we obtain
\beq
H_{\rm osc}^2 \;\simeq\; \frac{2V'(a_{\rm osc})}{ ca_{\rm osc}}.
\label{hubble}
\eeq
{ It should be noted that $V' /a> 0$ is always satisfied in our setup.}
For a given potential $V$,
 one can compute $\h_{\rm osc}$ by (\ref{starosc}) and (\ref{hubble}).
In the rest of this section we estimate the axion abundance for cases with a single cosine potential (\ref{single})
and with a flat-bottomed potential~(\ref{double}).

\subsection{{\bf Case with a single cosine potential}}
First let us consider a case in which the axion potential is given
by a single cosine potential  (\ref{single}). 
In this case, the potential (\ref{single}) can be  approximated by 
the quadratic potential (\ref{appquad}) unless the axion initially sits near the potential maximum.
We also assume that the axion mass is much smaller than the Hubble parameter during inflation.
Then, the axion starts to oscillate well after inflation
when the Hubble parameter becomes comparable to the axion mass.
According to (\ref{starosc}) and (\ref{hubble}), $\h_{\rm osc}$ is given by
\beq
\h_{\rm osc}\simeq\h_*\exp\left(-\frac{N_*m^2}{3H_{\rm inf}^2}-\frac{1}{4(c-3)}\right),\label{singleosc}
\eeq
and therefore, $\h_{\rm osc}$ is proportional to $\h_*$. After the axion starts to oscillate,
its energy density decreases like non-relativistic matter. If the axion starts to oscillate
in the radiation dominated era, the axion abundance is given by
\beq
\frac{\rho_a}{s} \;\simeq\; \frac{\frac{1}{2} m^2 f^2 
\theta_{\rm osc}^2}{\frac{2 \pi^2}{45} g_{*s}(T_{\rm osc}) T_{\rm osc}^3},
\eeq
where $g_{*s}(T)$ is the effective relativistic degrees of freedom contributing to the entropy density,
and $T_{\rm osc}$ is the plasma temperature given by
\begin{align}
T_{\rm osc} &= \lrfp{\pi^2 g_*(T_{\rm osc})}{90}{-\frac{1}{4}}\left(\frac{c}{2}\right)^{-\frac{1}{4}} \sqrt{m M_p},\\
&\simeq 0.2 {\rm \,MeV} \lrfp{g_*(T_{\rm osc})}{3.36}{-\frac{1}{4}} \lrfp{m}{10^{-17}{\rm\,eV}}{\frac{1}{2}}
\end{align}
with $M_p \simeq 2.4 \times 10^{18}$\,GeV being the reduced Planck mass, and
$g_*(T)$ is the effective relativistic degrees of freedom contributing to the energy density.
{ In the second equality, we have used $c=5$. As a result,} the axion density parameter is given by
\beq
\Omega_ah^2\simeq
0.2\, \h_{\rm osc}^2 \left(\frac{g_*(T_{\rm osc})}{3.36}\right)^{\frac{3}{4}}\left(\frac{g_{*s}(T_{\rm osc})}{3.91}\right)^{-1}\left(\frac{m}{10^{-17}\,{\rm eV}}\right)^{\frac{1}{2}}\left(\frac{f}{10^{16}\,{\rm GeV}}\right)^2,
\label{singleabundance}
\eeq
where $h\simeq0.7$ is the dimensionless Hubble parameter.
In order not to exceed the observed dark matter abundance,
 $\Omega_ch^2\simeq0.12$ \cite{Ade:2015xua},  for a general initial condition $\theta_{\rm osc} = {\cal O}(1)$ and the expected size
of the decay constant $f \sim \GEV{16}$, 
 the axion mass must be  as small as $m_a\lesssim10^{-17}$ eV.
Therefore, all the corresponding shift symmetry breaking terms must be extremely suppressed to explain the observed dark matter abundance. This places a certain constraint on the UV physics. 

\subsection{{\bf Case with a flat-bottomed potential}}
Now let us consider the case where the light axion mass is realized by the cancellation between two shift symmetry
breaking terms.
 In this case, the potential is approximated by (\ref{appquart}),
and the potential is almost quartic at the onset of oscillations for a general initial condition, $\theta_T \ll \theta_* \lesssim {\cal O}(1)$.
The curvature of the potential becomes equal to the Hubble parameter during inflation at 
\beq
\theta_H \;\equiv\;\frac{\sqrt{2} H_{\rm inf}}{\sqrt{\lambda} f} 
\simeq \frac{\sqrt{2} H_{\rm inf}}{\sqrt{n_2^2 - n_1^2} M_1}.
\eeq
Even if $\theta_{H} \lesssim {\cal O}(1) $ so that a part of the 
flat-bottomed potential (\ref{double}) has a curvature larger than
the Hubble rate, 
for $\theta \gtrsim \theta_H$,  the potential is so steep that the axion would soon move 
to the region $\theta \lesssim \theta_H$ during inflation. 
Thus, we assume $\theta_T \ll \theta_* \lesssim \theta_H$ in the following.

From (\ref{starosc}) and (\ref{hubble}),  $\h_{\rm osc}$ is given by
\begin{align}
\h_{\rm osc}
&\simeq \sqrt{\frac{2c-7}{2c-6}} \frac{\h_*}{\sqrt{1+\left(\frac{\h_*}{\h_c}\right)^2}},\label{osc}
\end{align}
with
\begin{align}
\h_c&\;\equiv\; \frac{3}{\sqrt{2 N_*}} \theta_{ H}.
\end{align}
Thus for $ \h_*  \lesssim  \h_c$, one sees that $\theta_{\rm osc}$ is
approximately proportional to $\theta_*$ as in the previous case.
On the other hand for $ \h_c \lesssim \h_* \lesssim \h_H$,
the dependence on~$\h_*$ is suppressed and $\h_{\rm osc}$ becomes
roughly $\h_c\sim\mathcal{O}(0.1)\,\h_H$.

\begin{figure}[t]
\centering
\includegraphics[width=10cm]{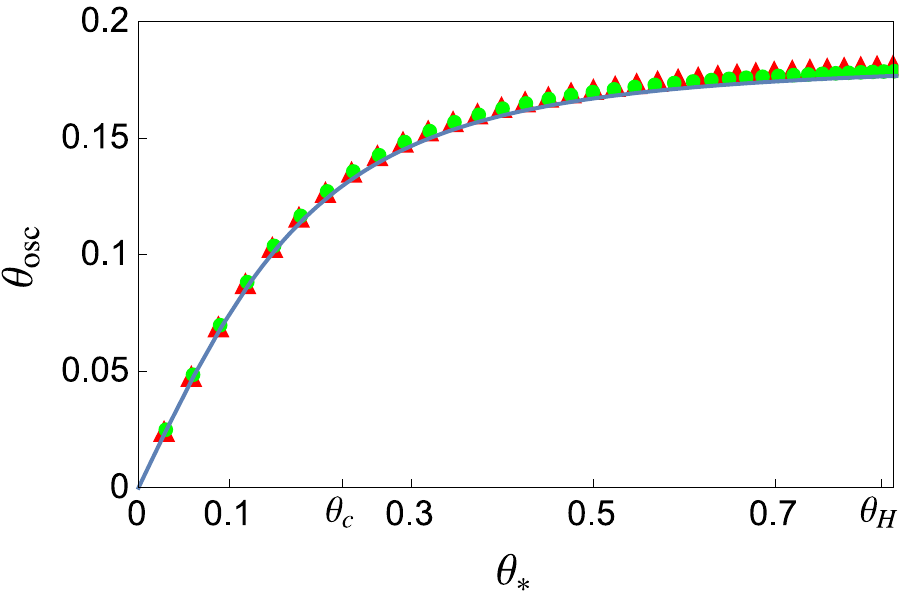}
  \caption{
  $\h_{\rm osc}$ as a function of $\h_*$ for $M_1/H_{\rm inf}=1$,
   $n_1=1$, $n_2=2$, and $N_*=60$. The solid (blue) line denotes the analytic result (\ref{osc}). The (red) triangles and (green) circles represent the numerical results for the full potential (\ref{double}) and  the quartic potential (\ref{appquart}), respectively.  See the text for details.
 }
\label{OSC}
\end{figure}

In Fig.~\ref{OSC} we show $\h_{\rm osc}$ as a function of $\h_*$ where
we have fixed the other parameters as $n_1=1$, $n_2=2$, $N_*=60$,
$M_1=H_{\rm inf}=10^{12}\,{\rm GeV}$, and $f=10^{16}\,{\rm GeV}$.
For the parameters adopted, $\theta_c \simeq 0.22$ and $\theta_H \simeq 0.82$.
The solid (blue) line denotes the analytic result (\ref{osc}) with $m=0$ and $c=9/2${, which implies that the axion starts to oscillate before reheating. The (red) triangles and (green) circles show numerical results obtained by solving the equation of motion  for the full potential (\ref{double}) and  the quartic potential (\ref{appquart}), respectively. 
} 
Here we note that $\h_{\rm osc}$ is insensitive to the value of~$m$, or
the difference between $M_1$ and $M_2$, as long as it is taken to be
sufficiently small.
One can see that the quartic approximation of the potential is fairly good for the parameters adopted. In the region where $\h_*>\h_c$, 
 $\h_{\rm osc}$ becomes less sensitive to the initial position
 $\theta_*$, and approaches a constant value of about
$0.18$.\footnote{We have adopted $\kappa = 1/2$ for the
definition of the onset of the oscillation~(\ref{onset}).
When using different values for $\kappa$ (e.g. $\kappa = 1$), the analytically
computed~$\h_{\rm osc}$ can deviate from the numerical one by a factor
of order unity, however the overall behavior of~$\h_{\rm osc}$ is
still captured by the analytic computation.
Hence changing the value of~$\kappa$ by an order-unity factor
only gives minor corrections to our discussions below.}

Next let us estimate the final axion abundance. 
The axion energy density decreases like radiation after it starts to oscillate until its oscillation
amplitude becomes smaller than $\theta_T$.
Let us focus on cases where
reheating takes place after the axion starts to oscillate, but before the axion
starts to behave like non-relativistic matter.
Then, bearing in mind that the total energy density of the background
universe~$\rho_{\mathrm{tot}}$ is (effectively) matter-dominated from
the end of inflation until reheating, 
the present axion-energy-density-to-entropy ratio is computed as
\begin{align}
\left.\frac{\rho_a}{s}\right|_0 
&= \left.\frac{\rho_a}{s}\right|_T\\
&=\rho_{a,T}^\frac{1}{4} \left.\frac{ \rho_{a}^\frac{3}{4}}{\rho_{\rm tot}}\right|_{\rm osc} \left.\frac{\rho_{\rm tot}}{s}\right|_{\rm reh}\\
&= \frac{3^\frac{11}{4} \theta_{\rm osc} f m T_{\rm reh}}{2^\frac{11}{2} \sqrt{\lambda} M_p^2}
\end{align}
where the subscripts $T$ 
means that the variable is evaluated when the oscillation amplitude of the
axion becomes equal to $\theta_T f$, and we have neglected the difference between $g_*(T_{\rm reh})$ and $g_{s*}(T_{\rm reh})$ since the reheating temperature is sufficiently high.
In the last equality, we have used $c = 9/2$ for
computing~$H_{\mathrm{osc}}$,  
and also used the following relations,
\begin{flalign}
 \r_{a,T}&\simeq \frac{9m^4}{2\l},\\
 \r_{a,{\rm osc}}&\simeq\frac{\l f^4 \theta_{\rm osc}^4}{4!}.
 \end{flalign}
The reheating temperature $T_{\rm reh}$ is defined by
\beq
T_{\rm reh} \;\equiv\;\lrfp{\pi^2 g_*}{90}{-\frac{1}{4}} \sqrt{\Gamma_{\rm inf} M_p},
\eeq
where $\Gamma_{\rm inf}$ is the inflaton decay rate and we have assumed an instant 
reheating at $H = \Gamma_{\rm inf}$.
 In terms of the density parameter, the axion abundance is given by
\begin{align}
\Omega_ah^2
& \simeq \frac{0.3}{n_2^2-n_1^2} \lrf{\theta_{\rm osc}}{0.1\theta_H} 
\left(\frac{m}{100\,{\rm MeV}}\right)
\left(\frac{f}{10^{16}\,{\rm GeV}}\right)^2
\left(\frac{T_{\rm reh}}{10^{10}\,{\rm GeV}}\right)
\non \\
&~~~~~~~~~\times
\left(\frac{M_1}{10^{12}\,{\rm GeV}}\right)^{-2}\left(\frac{H_{\rm inf}}{10^{12}\,{\rm GeV}}\right).
\label{abundance}
\end{align}
Therefore, the axion mass can be much heavier than the case of a single cosine term, and interestingly,
it happens to be close to the upper bound (\ref{uppm}) when the other 
parameters take the reference values shown in the parentheses. 
Such decaying dark matter with lifetime close to the current lower bound 
can partially relax the tension of $\sigma_8$~\cite{Enqvist:2015ara}.

\section{Axionic Isocurvature Perturbations}
\label{isocurvature}
During inflation the axion acquires quantum fluctuations which result in axionic isocurvature perturbations~\cite{Steinhardt:1983ia,Axenides:1983hj,Linde:1985yf,Seckel:1985tj}.
The current upper bound on the scale-invariant and uncorrelated isocurvature perturbation is \cite{Ade:2015lrj}
\beq
\mathcal{P}_S\; \lesssim\; 0.04 \,\mathcal{P}_R,
\eeq
where $\mathcal{P}_R (\simeq2.2\times10^{-9})$ and $\mathcal{P}_S$ 
are respectively the power spectra of curvature and isocurvature perturbations
on the pivot scale $k_{p} = 0.05 \, \mathrm{Mpc}^{-1}$.
The axionic isocurvature perturbation is given by~\cite{Kobayashi:2013nva}
\beq
\mathcal{P}_S=(r_a\,\Delta_{a,S})^2,
\eeq
\beq
\Delta_{a,S}=\frac{\partial \ln\Omega_{a}}{\partial \h_*}\frac{\delta a_{k_p}}{f},
\label{Ds}
\eeq
where $r_a$ is the fraction of the axion dark matter to the total dark matter, 
and  $\delta a_{k_p}$ denotes the axion fluctuation on the pivot scale
at the horizon exit. For $\sqrt{|V''(a_*)|}<3H_{\rm inf}/2$, 
and if the time-dependence of $V''(a)$ during inflation can be neglected,
the size of the fluctuation is
\beq
\delta a_{k_p}\simeq \beta(\nu) \frac{H_{\rm inf}}{2\pi} \label{deltaakp}
\eeq
with
\begin{align}
\beta(\nu) & \equiv \left(\frac{2\Gamma(|\nu|)}{\sqrt{\pi}}\right)\left(\frac{1}{2}\right)^{\frac{3}{2}-|\nu|},\\
\nu^2 & \equiv\frac{9}{4}-\frac{V''(a_*)}{H_{\rm inf}^2},
\end{align}
where $\Gamma(\nu)$ denotes the gamma function. In the limit of $|V''(a_*)|\ll H_{\rm inf}^2$, $\delta a_{k_p}$ reproduces
the well-known result, $H_{\rm inf}/(2\pi)$. On the other hand,
$\delta a_{k_p}$ can be suppressed if the curvature is comparable
to the Hubble parameter. For instance, if $V''(a_*) =  H_{\rm inf}^2$, $\beta(\nu)$  is about $0.82$.
(However we should also remark that the expression~(\ref{deltaakp}) is derived assuming
a constant~$V''$ during inflation.
Hence if $|V''|$ is as large as $H_{\mathrm{inf}}^2$, the axion
rapidly rolls during inflation and thus the time-dependence of~$V''$ may
become relevant, giving corrections to the evaluation of~$\delta a_{k_p}$.
For simplicity, here we ignore such effects.)

In the following we estimate the axionic isocurvature perturbations
for cases with a single cosine potential (\ref{single}) and a flat-bottomed potential (\ref{double}).
 
\subsection{{\bf Case with a single cosine potential}}
From (\ref{singleosc}) and (\ref{singleabundance}),  $\Omega_a$ is proportional to $\h_*^2$, and one obtains
\begin{align}
\Delta_{a,S}^2&=\left(\frac{H_{\rm inf}}{\pi f \h_*}\right)^2\\
&\simeq 1\times10^{-9}\frac{1}{\h_*^2}\left(\frac{H_{\rm inf}}{10^{12}\,{\rm GeV}}\right)^2\left(\frac{f}{10^{16}\,{\rm GeV}}\right)^{-2},\label{singleiso}
\end{align}
where we have used $\delta a_{k_p}=H_{\rm inf}/(2\pi)$ because the axion
mass $m$ ($\lesssim10^{-17}\,{\rm eV}$ for $f \sim 10^{16}\, \mathrm{GeV}$) is much smaller than $H_{\rm
inf}$.
We neglected the $\theta_*$-dependence of the relativistic degrees
of freedom $g_{*}(T_{\mathrm{osc}})$ and $g_{*s}(T_{\mathrm{osc}})$. 
For the reference values in the parentheses
with $\theta_* = {\cal O}(1)$ and $r_a = 1$,
the resultant axionic isocurvature perturbation 
${\cal P}_S$ exceeds the current bound by one order of magnitude. Therefore, $H_{\inf} \lesssim {\cal O}(10^{11})$\,GeV is needed for $f = \GEV{16}$ and $\theta_* = {\cal O}(1)$.

\subsection{{\bf Case with a flat-bottomed potential}}
Using (\ref{osc}) and (\ref{abundance}), we obtain
\begin{align}
\Delta_{a,S}^2&=\left(\left(\frac{1}{\h_*}-\frac{\h_*/\h_c^2}{1+\left(\h_*/\h_c\right)^2}\right)\beta(\nu)\frac{H_{\rm inf}}{2\pi f}
\right)^2\non \\
&\simeq 3\times10^{-10}\frac{1}{\h_*^2}
\left(\frac{\beta(\nu)}{1+(\h_*/\h_c)^2}\right)^2
\left(\frac{H_{\rm inf}}{10^{12}\,{\rm GeV}}\right)^2\left(\frac{f}{10^{16}\,{\rm GeV}}\right)^{-2},
\label{iso}
\end{align}
which is suppressed by a factor of
\beq
\frac{1}{4} \left(\frac{\beta(\nu)}{1+(\h_*/\h_c)^2}\right)^2
\eeq
compared to (\ref{singleiso}). 
The suppression is due to the following three factors. First,  the axion density parameter  $\Omega_a$  is proportional to $\h_{\rm osc}$ 
rather than $\theta_{\rm osc}^2$ (cf. Eqs.~(\ref{singleabundance}) and  (\ref{abundance})), which reduces
$\Delta_{a,S}^2$ by a factor of $4$ (see Eq. (\ref{Ds})). Secondly, $\theta_{\rm osc}$ is less sensitive to $\theta_*$, 
when $\theta_*$  is larger than $\theta_c$. This effect appears in the $\theta_*$ dependence of the denominator. 
Lastly, the axion fluctuation can be further suppressed by
$\beta({\nu})$, depending on the value of~$\nu$. 
In total, the axionic isocurvature perturbation can be suppressed by a few orders of magnitude compared
to the single cosine potential.

\begin{figure}[t]
\centering
\includegraphics[width=10cm]{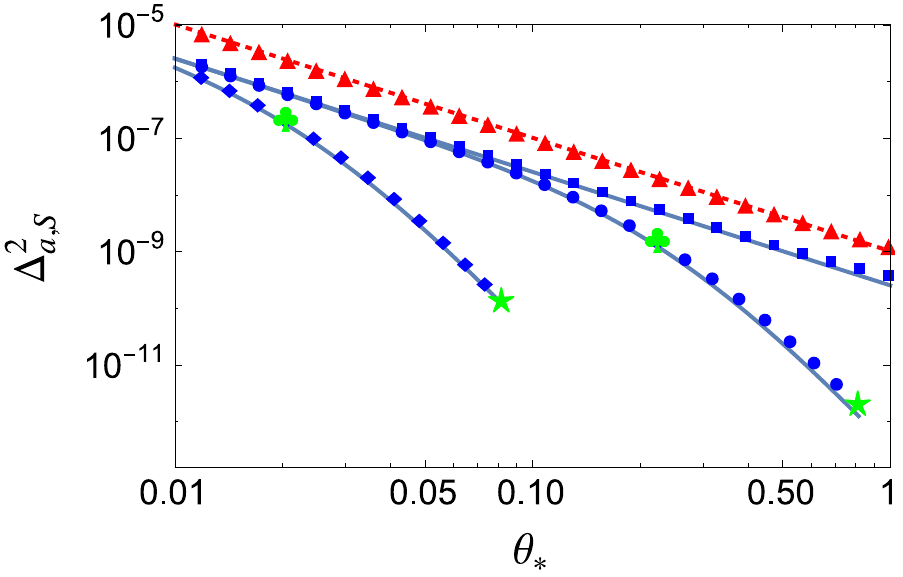}
\caption{Axionic isocurvature perturbation 
as a function of $\h_*$. We have fixed $n_1=1$, $n_2=2$, $N_*=60$, $H_{\rm inf}=10^{12}$ GeV, and $f=10^{16}$ GeV. The dotted (red) line denotes the
 analytic result for the single cosine potential. The solid (blue) lines denote analytic results for the flat-bottomed potential with $M_1<10^{11}$ GeV, $M_1=10^{12}$ GeV and $M_1=10^{13}$ GeV in order from the top.  
 The (red) triangles, the (blue) squares, circles, and diamonds represent the numerical results.}
\label{iso1}
\end{figure}

In Fig.~\ref{iso1} we show the axionic isocurvature perturbation, 
$\D_{a,S}^2$ as a function of $\h_*$. 
 We have fixed the inflationary parameters $H_{\rm inf}=10^{12}$ GeV, $N_*=60$. Other parameters are $f=10^{16}$ GeV, $n_1=1$ and $n_2=2$. The dotted (red) line denotes the
 analytic result for the single cosine potential. The solid (blue) lines denote analytic results for the flat-bottomed potential with $M_1<10^{11}$ GeV, $M_1=10^{12}$ GeV and $M_1=10^{13}$ GeV in order from the top.  
 We have also numerically computed the isocurvature perturbation using
 the $\delta \cal{N}$ formalism~\cite{Starobinsky:1986fxa,Sasaki:1995aw,Wands:2000dp,Lyth:2004gb}. 
 For a given $\h_*$, we have followed the evolution of the axion field from before the horizon exit until the oscillation amplitude becomes much 
smaller than $\h_T$, and  then evaluated the number of e-foldings
between a flat slice at the horizon exit and
a slice of uniform axion density after the axion have
started to behave as non-relativistic matter.  We have estimated the factor $\b(\nu)$ by using the numerical value of $V''$ at horizon exit.
 The (red) triangles represent the numerical result for the single cosine potential with $m=10^9\,{\rm GeV}$.\footnote{{ Note that we have used 
 a rather large $m$ just for 
 efficiency of the numerical calculation. However, the value of $m$ is
 irrelevant here because $\Delta_{a,S}^2$ is independent of $m$.}
We also remark that $\Delta_{a,S}^2$ is insensitive to
the precise value of the reheating scale;
for the numerical computations
we have chosen the reheating scale such that reheating
takes place after the onset of the axion oscillation, but before the
axion starts to behave as non-relativistic matter.} 
 The (blue) squares, circles, and diamonds represent the numerical
 result for the flat-bottomed potential with $M_1=10^{11}$ GeV,
 $M_1=10^{12}$ GeV, and $M_1=10^{13}$ GeV, respectively. Here, we have
 fixed to $m=10^{-3}M_1$. The (green) stars denote $\h_H$ and (green)  clubs denote $\h_c$.
One can see from the figure 
 that the analytic and numerical results agree well with each other, and that
 the isocurvature perturbations of the  flat-bottomed potential is at
 most three orders of magnitude smaller than that of the single cosine potential 
 around $\h_*\sim\h_H$. In the region $\h_*>\h_H$,  the isocurvature perturbation is significantly suppressed
 because the axion is heavier than the Hubble parameter during
 inflation, and thus the axion fluctuation becomes highly
 suppressed. However, we also note that such a 
 heavy axion is considered to move quickly to smaller field values, and
 thus $\theta_* \gtrsim \theta_H$ would not be a plausible choice of the initial condition. 

In summary, the axionic isocurvature perturbations can be suppressed by a few orders of magnitude in a case with the flat-bottomed potential,
compared to the ordinary cosine potential. As a result, if the axion accounts for the observed dark matter density, the isocurvature constraint
on the inflation scale, $H_{\rm inf}$, is relaxed by one order of magnitude.

\section{Conclusions}

So far we have focused on the axion dynamics with a flat-bottomed potential, but the same argument
can be straightforwardly applied to other fields such as the Polonyi field, and open string and
complex structure moduli fields. In general, the abundance of such scalars with the flat-bottomed 
potential is significantly suppressed, which ameliorates cosmological constraints and enables them
to be a viable candidate for dark matter.

If there are many axions in nature as suggested by the Axiverse or axion landscape, 
some of them may play important cosmological roles such as driving
inflation, baryogenesis, and serving as dark matter.
The purpose of this paper is to study a possibility that the axion is cosmologically stable 
and contributes to dark matter.

In order to account for dark matter, the axion's lifetime must be longer than the present age of the Universe. 
This requires the axion mass  to be sufficiently light.  Broadly speaking,
there are two ways to realize the light axion mass. One is to suppress all the terms in the potential.
In this case, the axion potential can be approximated by a single cosine term which dominates the potential,
and the axion mass must be extremely small ($m \lesssim 10^{-17}$\,eV for
a decay constant of $f \sim \GEV{16}$)
 to give the right abundance of dark matter. For such extremely small mass,
the stability of the axion is automatically satisfied. 

 The other is that multiple terms conspire to make the axion mass light at the potential minimum.
 This  is possible if the axion potential receives multiple contributions as suggested by  the axion landscape.
The axion potential in this case has a flat-bottomed shape, which can be
 approximated by a quartic potential plus a relatively tiny mass term. 
We have found that  the axion abundance is highly suppressed since the axion behaves as radiation in the early Universe,
and as a result, the axion mass can be rather large (e.g. $m \sim 100$\,MeV). For such a relatively heavy axion mass,
 the axion's lifetime  may not be very different from the present age of the Universe. In particular,
if such relatively heavy axion decays into e.g. hidden photons
with a lifetime close to the current lower limit, it can partially relax the tension of $\sigma_8$. 
  We have also shown that the axionic isocurvature perturbations can be suppressed by a few orders of magnitude compared
  to the case of a single cosine potential  (see Fig.~\ref{iso1}), which relaxes the isocurvature constraint on the
  inflationary scale. 
  
In summary, the axion dark matter with a flat-bottomed potential allows a relatively heavy axion mass, and
its abundance as well as isocurvature perturbation can be significantly suppressed compared to the ordinary
single cosine potential. Such decaying axion dark matter will be either favored or severely constrained by the future
observations of CMB and large-scale structure.

\section*{Acknowledgment}

T.K. would like to thank the Particle Theory and Cosmology Group of
Tohoku University for hospitality during the initiation of this work. 
This work is supported by
Tohoku University Division for Interdisciplinary Advanced Research and Education (R.D.),
INFN INDARK PD51 grant (T.K.), 
MEXT KAKENHI Grant Numbers 15H05889 and 15K21733 (F.T.),
JSPS KAKENHI Grant Numbers  26247042 and 26287039 (F.T.), and
World Premier International Research Center Initiative (WPI Initiative), MEXT, Japan~(F.T.).

\end{document}